\documentclass[epj]{svjour}
\usepackage{epsfig}
\newcommand{\beq}{\begin{equation}}
\newcommand{\eeq}{\end{equation}}
\newcommand{\bdis}{\begin{displaymath}}
\newcommand{\edis}{\end{displaymath}}
\newcommand{\bea}{\begin{eqnarray}}
\newcommand{\eea}{\end{eqnarray}}
\newcommand{\barr}{\begin{array}}
\newcommand{\earr}{\end{array}}

\begin{document}

\title{Damage in fiber bundle models}

\author{Ferenc Kun\inst{1,2}\and Stefano Zapperi\inst{3,4} 
\and Hans. J. Herrmann\inst{1,3}} 

\institute{ICA 1, University of Stuttgart, 
Pfaffenwaldring 27, 70569 Stuttgart, Germany
\and Department of Theoretical Physics, University of Debrecen, 
P.O.Box: 5, H-4010 Debrecen, Hungary,
\and PMMH-ESPCI, 10 Rue Vauquelin, 75231 Paris CEDEX 05, France,\and
INFM sezione di Roma 1, Universit\`a "La Sapienza", 
P.le A. Moro 2 00185 Roma, Italy.}

\date{\today}
\abstract{
We introduce a continuous damage  fiber bundle model 
and compare its behavior with that of dry fiber
bundles. Several interesting constitutive behaviors, such as
plasticity, are found in this model depending 
on the value of the damage parameter 
and on the form of the disorder distribution.
We compare the constitutive behavior of global load transfer models,
obtained analytically, with local load transfer models 
numerical simulations. The evolution of the damage is
studied analyzing the cluster statistics for dry and
continous damage fiber bundles. Finally, it is shown
that quenched random thresholds enhance damage localization.
}
\maketitle

\section{introduction}

The rupture of disordered media has recently attracted much technological
and industrial interest and has been widely
studied in statistical physics. It has been suggested by several authors
that the failure of a disordered material subjected to 
an increasing external load shares many features 
with thermodynamic phase transitions.
In particular, a stressed solid can be considered to be
in a metastable state \cite{buse}
and the point of global failure can be seen as
a nucleation process in a first order transition near a spinodal
\cite{rk,sel1,sel2}. Thus, the power law behavior observed
experimentally in the acoustic emission before failure
\cite{ciliberto,strauven,ae} has been compared  with the 
mean-field scaling expected close to a spinodal point \cite{zrsv}.
In analogy with spinodal nucleation \cite{sn}, 
scaling behavior can only be seen when long-range interactions
are present, as it is the case for elasticity, but should not be
observable when the stress transfer function is short ranged.
This observation is confirmed in fracture models with short-range 
elastic forces, which usually do not show scaling 
\cite{hansen2,hansen3}.

Most of the theoretical investigations in this field rely on large
scale computer simulation of lattice models where the elastic medium
is represented by a spring (beam) network, and disorder is captured
either by random dilution or by assigning random failure thresholds to
the bonds \cite{hans}. The failure rule usually applied in lattice
models is discontinuous and irreversible: when the local load 
exceeds the failure threshold of a bond, 
the bond is removed from the calculations ({\em i.e.} its
elastic modulus is set to $0$). Furthermore, failed bonds
are never restored (no healing). 
Very recently, a novel continuous damage law has been introduced 
in lattice models \cite{zvs}. In the framework of
this model when the failure threshold of a bond is exceeded the
elastic modulus of the bond is reduced by a factor $a$ ($0 < a<1$), 
furthermore, multiple failures of bonds are allowed. This 
description of damage in terms of a continuous parameter corresponds to
consider the system at a length scale larger than the typical crack 
size. Computer simulations have revealed some remarkable 
features of the model: after some transients the
system tends to a steady state which is macroscopically plastic, and is
characterized by a power law distributed avalanches of breaking events. 

A very important class of models of material failure 
are the fiber bundle models (FBM) \cite{daniels,coleman,krajcinovic,sornette1,sornette2,moukarzel,phoenix1,phoenix2,phoenix3,phoenix4,beyerlein,curtin1,curtin2,curtin3,hild,curtin4,curtin5,curtin6}, 
which have been extensively
studied during the past years. These models consists of a set of
parallel fibers having statistically distributed strength. 
The sample is loaded parallel to the fibers direction, and the fibers 
fail if the load on them exceeds their threshold value. In stress 
controlled experiments, after each fiber failure the load carried by the 
broken fiber is redistributed among the intact ones. Among the several
theoretical approaches, one simplification that makes the problem
analytically tractable is the assumption of global load transfer,
which means that after each fiber breaking the stress is equally
distributed on the intact fibers neglecting stress enhancement in the
vicinity of failed regions \cite{daniels,coleman,krajcinovic,sornette1,sornette2,moukarzel,phoenix1,phoenix2,phoenix3,phoenix4,curtin4,curtin5,curtin6}.
The relevance of FBM is manifold: in spite of their 
simplicity these models capture the most important aspects of
material damage and due to the analytic solutions they provide a deeper 
understanding of the fracture process.  
Furthermore, they serve as a basis for more realistic damage 
models having also practical importance. The very successful 
micromechanical models of fiber reinforced composites 
are improved variants of FBM
taking into account stress localization (local load transfer)
\cite{moukarzel,phoenix1,phoenix2,beyerlein}, 
the effect of matrix material between fibers
\cite{phoenix1,phoenix2,phoenix3,phoenix4,curtin4,curtin5,curtin6},
and possible non-linear behavior of fibers \cite{krajcinovic}.  
Previous studies of FBM addressed the macroscopic 
constitutive behavior, the reliability and size scaling of the
global material strength, and the avalanches of 
fiber breaks preceding ultimate failure 
\cite{hansen2,hansen3,hansen1,newman1,newman2,moreno,rava}.

In this paper we generalize the FBM
applying a continuous damage law for the elements in the spirit of 
Ref.~\cite{zvs}. Emphasis is put on the microstructure of damage 
and its evolution with increasing load.
For the case of global load transfer in the continuous damage model, 
we derive an exact analytic expression for the constitutive 
behavior and show that the system reaches asymptotically a 
steady state, which is macroscopically plastic. 
It is demonstrated that the continuous damage model provides a broad 
spectrum of description of materials by varying its parameters 
and for special choices of the parameter values the model recovers 
the dry FBM and other micromechanical models of composites 
known in the literature.
Next, we present a theoretical investigation of damaging in
FBM by studying the 'dry' FBM  varying 
the range of load transfer. In the case of global load transfer
the model approaches the failure point by scaling laws, analogous
to those observed close to a spinodal \cite{zrsv}. However, scaling
is not observed for local load transfer FBM \cite{hansen2,hansen3}
as it is expected for a spinodal instability, 
which can only be observed in mean-field
theory. Increasing the range of interactions, one can observe that
the spinodal point, defined in the global load transfer FBM, 
is approached. The evolution of damage is compared in the local load
transfer dry and continuous damage FBM.  Finally, we analyze the
effect of quenched random threshold on damage localization.

The paper is organized as follows: in section 2 we describe the
FBM and derive their constitutive behavior in section 3. 
In section 4 we discuss the local load transfer FMB focusing on the
constitutive behaviour and on the cluster analysis, and in
section 5 we explore the role of the type of disorder in the evolution 
of damage. Section 6 is devoted to discussion and conclusions.

\section{Models}

\begin{figure}
\begin{center}
\epsfig{bbllx=165,bblly=305,bburx=432,bbury=410,
file=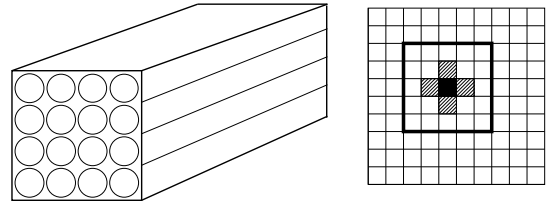,
  width=8cm}
\caption{The geometry of the FBM. The uniaxial
  fiber bundle (left) is modeled on a square lattice (right)
  corresponding to a cross section of the specimen. The black
  plaquette indicates a broken fiber, its nearest neighborhood is
  shadowed, and the bold line shows the neighborhood for the range
  of interaction $R=2$.
} 
\label{fig:1}\end{center}
\end{figure}

The system under consideration is composed of $N$ 
fibers assembled in parallel on a two dimensional square lattice of 
side length $L$, {\em i.e.} $N=L^2$. 
The geometrical structure of the model is illustrated in 
Fig.\ \ref{fig:1}. The square lattice corresponds to a cross section
of a unidirectional fiber ensemble. 
In FBM, the fibers are considered to be linearly elastic until
breaking (brittle failure)
with identical Young-modulus $E_f$ but with random
failure thresholds $d_i$, $i= 1, \ldots , N$. The failure strength
$d_i$ of individual fibers is supposed to be independent identically
distributed random variables with a cumulative probability distribution
$P(d)$. The fiber bundle is supposed to be loaded uniaxially, 
and load $F$ applied parallel to the fibers 
gives rise to a strain $f$ of the bundle. 
When a fiber experiences a local load larger than its failure
threshold the fiber fails. In dry FBM there is no matrix material
present, which implies that broken fibers do not support load any
more, and their load is redistributed to the surviving fibers.

In the global load transfer FBM, after failure the load
is transfered equally to all the remaining intact fibers, so that
the load on fiber $i$ is simply given by $F_i = F/n_s(F)$, where 
$n_s(F)$ is the 
total number of surviving fibers for a load $F$. This also 
implies that the range of interaction between fibers is
infinite, and hence, the global load transfer corresponds to the
mean field treatment of FBM.
In the local load transfer FBM the load
is transfered equally only to the surviving nearest neighbor fibers,
giving rise to high stress concentration around failed regions (see
also Fig.\ \ref{fig:1}).
We also study intermediate situations in which the load
is transfered to a local neighborhood surrounding the failed
fiber ({\em i.e.} a  square of radius
$R$ centered on the failed fiber, see Fig.~\ref{fig:1}). 
This model interpolates between the nearest neighbor
local FBM and the global FBM as the range of interaction  
is increased. 

Next we generalize the model replacing the brittle failure
of fibers by a continuous damage parameter \cite{zvs}. 
When the load on a fiber reaches the threshold value $d_i$ 
the stiffness $E_f$ of the fiber is reduced by a factor $0 < a < 1$. 
The characterization of damage by  a continuous parameter corresponds to 
describe the system on length scales larger than the typical crack size.
This can be interpreted such that the smallest elements of the model 
are fibers and the continuous damage is due to cracking inside fibers.
However, the model can also be considered as the discretization of 
the system on length scales larger than the size of single fibers, so that
one element of the model consists of a collection of fibers with matrix 
material in between. In this case the microscopic damage mechanism 
resulting in multiple failure of the elements is the gradual cracking of 
matrix and the breaking of fibers.

\begin{figure}
\begin{center}
\epsfig{bbllx=0,bblly=-15,bburx=382,bbury=460,file=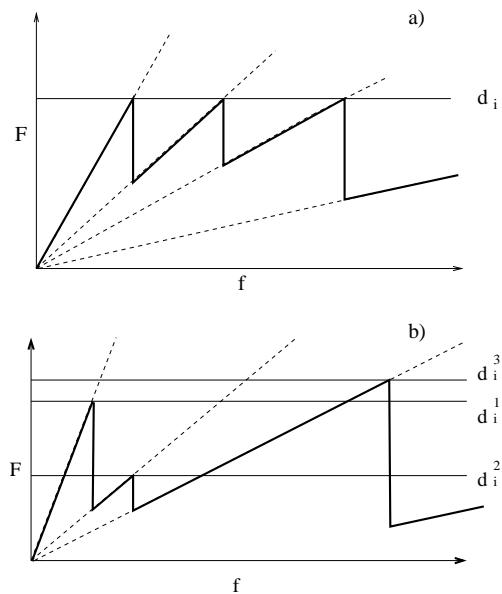,
  width=6cm}
\caption{The constitutive behavior of a single fiber of the
  continuous damage model when multiple failure is allowed.
(a) Quenched disorder: the
 horizontal line indicates the damage threshold $d_i$, which is
 constant in time for each fiber.
(b) Annealed disorder: a new threshold is extracted at random
after each failure.}
\label{fig:2}\end{center}
\end{figure}

In the following the elements of the continuous damage FBM
will be called fibers, but we have the above two possible
interpretations in mind. 
Once the fiber $i$ has failed its load is reduced to  
$af_i$ and the rest of the load $(1-a)f_i$ is distributed
equally among all the other fibers (global stress transfer)
or among the neighboring fibers (local stress transfer).
In principle, a fiber can now fail more than once and we define
$k_{max}$ as the maximum number of failures allowed per fiber.
We will first study the model for finite $k_{max}$ and eventually
take the limit $k_{max}\to\infty$. It is important to note that
once a fiber has failed, we can either keep the same
failure threshold (quenched disorder) or chose a different
one of the same distribution (annealed disorder), which can model
microscopic rearrangements in the material. The failure rules  
of the model in the two cases are illustrated in Fig.\ \ref{fig:2}.
In the following sections we will analyze both cases, showing
that there are differences in the microstructure of damage between quenched
and annealed disorder in this problem.

\section{Constitutive laws}

Here we derive the constitutive law for continuous damage
FBM and show how the FBMs used in the literature can be recovered in 
particular limits. We first consider the case 
in which fibers are allowed to fail only once: the
constitutive equation reads as
\begin{eqnarray}
  \label{eq:constone}
  \frac{F}{N} = f(1-P(f)) + af P(f),
\end{eqnarray}
where $P(f)$ and $1-P(f)$ are the fraction of failed and intact
fibers, respectively, and the Young-modulus $E_f$ of intact fibers is taken
to be unity. In Eq.\ (\ref{eq:constone}) the first term provides the
load carried by intact fibers while the second term is the contribution
of the failed ones. Note that this particular case together with the
parameter choice $a=0$ ({\em i.e.} broken fibers carry no load) 
corresponds to the dry FBM  
\cite{daniels,coleman,sornette1,sornette2}, 
while setting 
$a=0.5$ in Eq.\ (\ref{eq:constone}) we recover
the so-called micromechanical model of
fiber reinforced ceramic matrix composites (CMC's), which has been 
extensively studied in the literature
\cite{curtin1,curtin2,curtin3,hild}. In CMC's the
 physical origin of the load bearing capacity of failed fibers is that
in the vicinity of the broken face of the fiber 
the fiber-matrix interface debonds
and the stress builds up again in the failed fiber through the sliding
fiber-matrix interface.

When the fibers are allowed to fail more than once we have to
distinguish  between quenched and annealed disorder.  

{\it (i) Quenched disorder:} 
When the fibers are allowed to fail twice the constitutive equation
can be written as

\begin{eqnarray}
  \label{eq:consttwo1}
  \frac{F}{N} = f(1-P(f)) + af \left[P(f)-P(af)\right] + a^2 f P(af),
\end{eqnarray}
where $\left[P(f)-P(af)\right]$ is the fraction of those fibers 
which failed
only once, and $P(af)$ provides the fraction of fibers which failed
already twice.
In the general case, when fibers are allowed to fail $k_{max}$
times, where $k_{max}$ can also go to infinity, the constitutive
equation can be cast into the form 
\begin{eqnarray}
  \label{eq:constk1}
  \frac{F}{N} =&& f(1-P(f)) + \sum_{i=1}^{k_{max}-1} a^i f \nonumber
  \left[P(a^{i-1}f)-P(a^if)\right] \\ 
  && + a^{k_{max}} fP(a^{k_{max}-1}f)
\end{eqnarray}

{\it(ii) Annealed disorder:}
As in the previous case we consider first the
case in which fibers are allowed to fail twice, obtaining
\begin{eqnarray}
  \label{eq:consttwo}
  \frac{F}{N} &=& f(1-P(f)) + af P(f)(1-P(af)) + \\
              & & + a^2 f P(f)P(af),  \nonumber
\end{eqnarray}
where $P(f)(1-P(af))$ is the fraction of fibers which failed
only once, and $P(f)P(af)$ is the fraction of fibers which failed
already twice. Finally, when fibers are allowed to fail $k_{max}$
times, where $k_{max}$ can also go to infinity, the constitutive
equation is given by
\begin{eqnarray}
  \label{eq:constk}
  \frac{F}{N} &=&  \sum_{i=0}^{k_{max}-1} a^if\left[1-P(a^if)
  \right]\prod_{j=0}^{i-1}P(a^jf) + \\
              & &  + a^{k_{max}} f\prod_{i=0}^{k_{max}-1}P(a^if).  \nonumber
\end{eqnarray}

\begin{figure}[htb]
\begin{center}
\epsfig{bbllx=125,bblly=75,bburx=467,bbury=686,
file=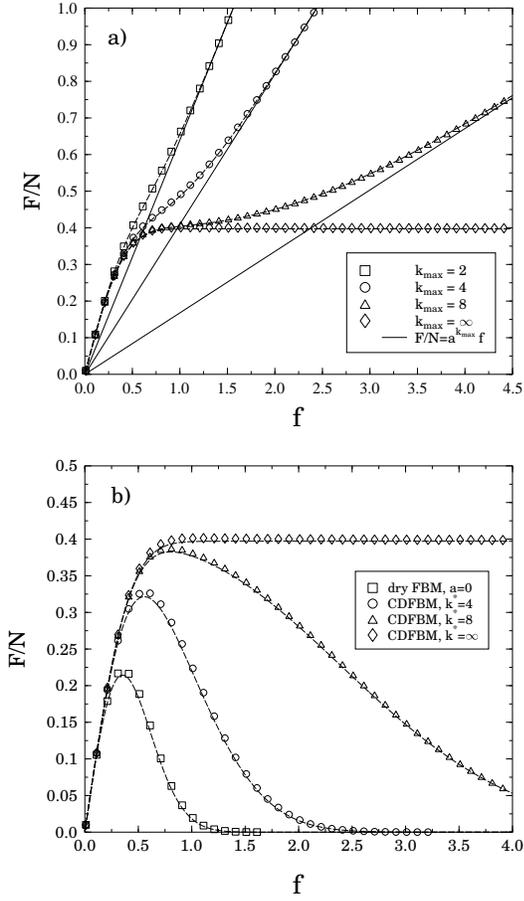,
  width=7cm}
\caption{(a) The constitutive law for the global stress transfer 
continuous damage FBM (CDFBM) for different values of $k_{max}$ for
$m=2.0$, $a=0.8$ and quenched disorder. Plastic
behavior is obtained in the limit of $k_{max}\to\infty$. 
(b) Comparison between dry and continuous damage FBM. If we allow for 
brittle failure after $k^*$ damage events, we obtain
a plastic plateau followed by brittle failure. Symbols refer
to simulations of bundles of size $N=128^2$ and lines to
the analytic calculations.}
\label{fig:31}\end{center}
\end{figure}

In Fig.\ \ref{fig:31} 
we show the explicit form of the constitutive
law for quenched disorder 
for different values of $k_{max}$ in the case of the Weibull distribution
\begin{equation}
P(d)=1-\exp(-(d/d_c)^{m}),
\end{equation}
where $m$ is the Weibull modulus and $d_c$ denotes the characteristic
strength of fibers. It is important to remark that the constitutive
laws derived above are exact only in the infinite size limit
($N\to\infty$), while fluctuations in the value of the failure stress
$F_c$ have been observed and studied for finite size bundles.
For this reason, we compare the theoretical results with
numerical simulations of bundles of size $N=128^2$. 
The agreement between simulations and theory
turns out to be satisfactory both for quenched (Fig.~\ref{fig:31})
and annealed disorder and reflects
the fact that for global load sharing finite size fluctuations,
for instance for $F_c$, should scale as $1/\sqrt(N)$. This behavior
has to be contrasted with local-load sharing fiber bundles
where finite size effects are very strong, as we will 
discuss in the following.

In Fig.\ \ref{fig:31}a 
the fibers are supposed to have $a^{k_{max}}$ 
residual stiffness after having failed $k_{max}$ times, which gives rise 
to hardening of the material, {\em i.e.} the $F/N$ curves asymptotically 
tend to straight lines with slope $a^{k_{max}}$. 
Increasing $k_{max}$ the hardening part of 
the constitutive behavior is preceded by a longer and longer 
plastic plateau, and in the limiting case of $k_{max}\to\infty$ the materials 
behavior becomes completely plastic (see Fig.\ \ref{fig:31}).
A similar plateau and asymptotic
linear behavior has been observed in brittle matrix composites, where
the multiple cracking of matrix turned to be responsible for the
relatively broad plateau of the constitutive behavior, and the
asymptotic linear part is due to the linear elastic behavior of
fibers remained intact after matrix cracking \cite{review}. 

In order to describe macroscopic cracking and global failure 
instead of hardening,
the residual stiffness of the fibers has to be set to zero 
after a maximum number $k^*$ of allowed failures \cite{zvs}. In this case
the constitutive law can be obtained from the general form 
Eqs.\ (\ref{eq:constk1}) and (\ref{eq:constk}) 
by skipping the last term corresponding to the 
residual stiffness of fibers, and by setting $k_{max} = k^*$ in the 
remaining part.  A comparison of the 
constitutive laws of the dry and continuous damage FBM is presented 
in Fig.\ \ref{fig:31}b for the case of quenched disorder. Annealed
disorder yields similar results. One can observe that the dry 
FBM constitutive law has a relatively sharp maximum, however, 
the continuous damage FBM curves exhibit a plateau whose length increases 
with increasing $k^*$. Note that  
the maximum value of $F/N$ corresponds to the macroscopic strength of the 
material and  
in stress controlled experiments the plateau and the decreasing part 
of the curves cannot be reached. 
However, by controlling the strain $f$, the plateau and the decreasing 
regime can also be realized. The value of the driving stress $\sigma
\equiv F/N$  corresponding to the plastic plateau is determined by
the damage parameter $a$, while the length of the plateau is
controlled by $k_{max}$ and $k^*$.

\begin{figure}[htb]
\begin{center}
\epsfig{bbllx=140,bblly=411,bburx=473,bbury=700,
file=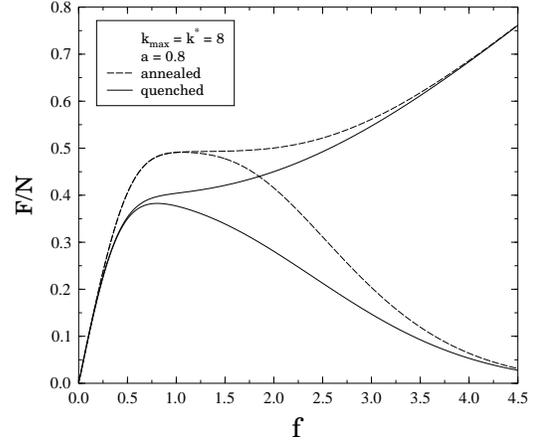,
  width=7cm}
\caption{Comparison of the constitutive laws of
the quenched and annealed case for $a=0.8$, 
$k_{max}=8$ and $k^*=8$.}
\label{fig:4}\end{center}
\end{figure}

In Fig.~\ref{fig:4}, we directly compare the constitutive 
law for quenched and annealed disorder and confirm that the
differences between the cases are very small. In particular,
all the basic constitutive behavior are reproduced in the
two cases.

It is important to remark that the behavior of the dry FBM model
($a=0$) under unloading and reloading to the original stress level is
completely linear, since no new damage can occur during
unloading-reloading sequences and the effect of the matrix material is
completely neglected.  
This also implies that in each damage state the model is completely 
characterized by the Young modulus defined as the slope of the unloading
curve. If the value of the damage parameter is larger than $0$ $(a >
0)$ the behavior of the system under unloading and reloading is
rather complicated. Due to the sliding of broken fibers with respect to
the matrix, hysteresis loops and remaining inelastic
strain occur (for examples see Ref.\ \cite{kun} and references
therein). 

\section{Local load transfer FBM}

\subsection{Constitutive behaviour}
To study the effect of stress enhancement around failed fibers on the
damage evolution and on the macroscopic constitutive behavior 
we employ local load transfer for the stress 
redistribution after fiber failure
\cite{phoenix1,curtin1,curtin2,curtin3}. Since
this case cannot be treated analytically,  we perform
numerical simulations in the dry FBM model:
after fiber failure the load is redistributed  
on the intact nearest neighbors of the failed fiber using periodic 
boundary condition on the square lattice (see also Fig.\ \ref{fig:1}).
For simplicity, in this case the strength of fibers $d_i$ has a
uniform distribution between $0$ and $1$.
The algorithm to simulate the loading process is 
as follows: $(i)$ we impose on all the fibers the same load, equal to
the smallest failure threshold, which results in breaking of the weakest
element. $(ii)$ The load carried by the failed fiber is redistributed 
on the intact nearest neighbors, and the load of the broken fiber 
is set to zero.
$(iii)$ After the stress redistribution, those fibers whose load exceeds their
failure threshold $d_i$ are identified and removed from the calculation, 
and the simulation is continued with point $(ii)$. If the
configuration obtained after 
the stress redistribution is stable, the global load is increased 
to cause the failure of one more fiber and the simulation is 
continued with point $(ii)$.
This procedure goes on until all fibers are broken.
The applied stress just before global failure is considered to be the 
failure strength of the model solid. Simulations were performed with
system sizes $L=16, 32, 64, 128$.

\begin{figure}[htb]
\begin{center}
\epsfig{bbllx=125,bblly=375,bburx=467,bbury=666,
file=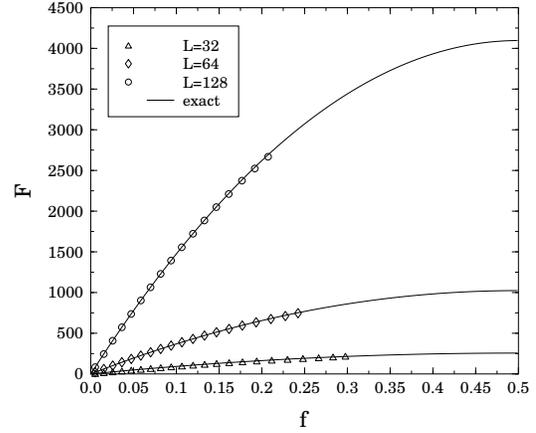,
  width=7cm}
\caption{The constitutive law for the global stress transfer 
dry FBM (solid line) is compared with the local stress transfer model.
Increasing the system size the failure stress decreases.}
\label{fig:5}\end{center}
\end{figure}

The constitutive behavior of the local and global load transfer dry FBM 
is compared in Fig.\ \ref{fig:5}.  For clarity, the total force $F$ 
(instead of stress) is presented as a function of strain $f$, 
for several different system sizes $L$. Note that $N=L^2$ is 
chosen the same for global and local load sharing simulations. 
In the case of global load transfer the $N\to\infty$
constitutive law can be obtained 
exactly by substituting the cumulative probability distribution
$P(f) = f$ of the uniform distribution into the general form Eq.\ 
(\ref{eq:constone}) and setting the damage parameter $a=0$:
\begin{eqnarray}
F = N f(1-P(f))=N f(1-f), \qquad   f \in [0,0.5],
\label{eq:constunif}
\end{eqnarray} 
the strain corresponding to macroscopic failure is $f_c=0.5$.
It can be seen in Fig.\ \ref{fig:5} that the macroscopic constitutive 
behavior for local load transfer always coincides with the 
global FBM solution, however, the macroscopic failure strength is 
substantially reduced in the local case giving rise to more brittle
constitutive behavior. It is interesting to note that increasing 
the system size $L$ the failure strength of the local FBM decreases,
showing the logarithmic size effect also found in the one dimensional 
local FBM \cite{hansen3} and in two dimensional
fuse networks \cite{duxbury}, while the global load 
transfer case does not have size dependence. 

\begin{figure}[htb]

\begin{center}
\epsfig{bbllx=125,bblly=375,bburx=467,bbury=666,
file=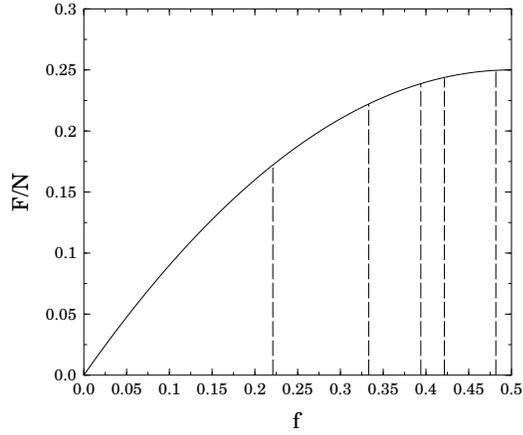,
  width=7cm}
\caption{The constitutive law for the global stress transfer 
dry FBM (solid line) is compared with the local stress transfer FBM
for different interaction ranges $R$.
The values of $R$ corresponding to the
consecutive vertical dashed lines are $1, 3, 5, 11, 15$ from left to
right, and the system size $L=128$ was chosen. Increasing the
interaction range the failure stress  
increases approaching the value predicted by the global 
stress transfer model. }  
\label{fig:6}\end{center}
\end{figure}

To get a deeper understanding of the behavior of the system as a function
of the range $R$ of load redistribution, we perform simulations 
by redistributing the load after fiber failure on the intact fibers in a 
square of side length $2R+1$ centered on the failed fiber. The range 
of load redistribution $R$ is varied between
$1$ and $(L-1)/2$.
Note that $R=1$ corresponds to local load transfer on
nearest and next-nearest neighbors, while the limiting case of $R=(L-1)/2$ 
recovers the 'infinite' range global load transfer. The comparison of
the constitutive behavior in the local and global load transfer case
is presented in Fig.\ \ref{fig:6}.
Simulations reveal that the constitutive 
laws obtained at different $R$ values always fall onto the curve 
of the global load transfer case
and the macroscopic failure strength increases with 
increasing range of interaction $R$ approaching the strength of global
FBM. For clarity, in Fig.\ \ref{fig:6} we indicated by
vertical dashed lines the position of global failure at different
values of $R$.

\begin{figure}[htb]

\begin{center}
\epsfig{bbllx=125,bblly=360,bburx=470,bbury=680,
file=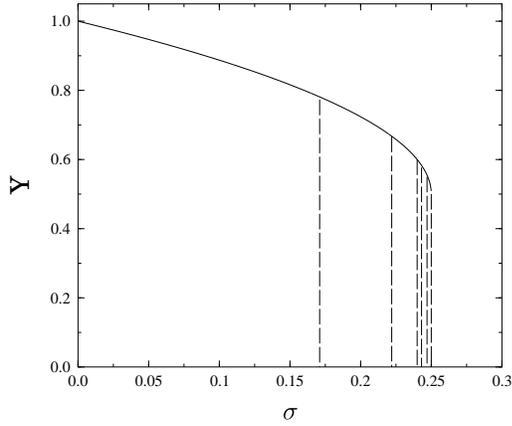,
  width=7cm}
\caption{The Young modulus for the global stress transfer 
dry FBM is compared with the local stress transfer model
for different interaction ranges $R$. The values of $R$ corresponding to the
vertical dashed lines are the same 
as in Fig.\ \protect\ref{fig:6}.}
\label{fig:7}\end{center}
\end{figure}

To characterize the elastic response of the dry FBM model in a given
damage state, 
we compute the Young modulus $Y$, defined as 
the slope of unloading curves as a 
function of the driving stress $\sigma \equiv F/N$:
\begin{eqnarray}
Y(\sigma)=\frac{ \sigma}{f(\sigma)} = 1-P(f(\sigma)) = 1-f(\sigma).
\end{eqnarray}
Using the constitutive law Eq.\ (\ref{eq:constunif}) for the global 
load transfer case, $Y$ can be 
written into a closed form as a function of stress
\begin{eqnarray}
Y(\sigma)=\frac12 \left[1+ \sqrt{1-4\sigma} \right].
\end{eqnarray}
The results on $Y$ for global and local load transfer are 
shown in Fig.\ \ref{fig:7}, where the vertical dashed lines 
indicate the position of macroscopic failure at different values
of the redistribution range $R$. 
It can be seen that at the failure point $Y(\sigma)$ has a discrete
jump, the size of which decreases with increasing $R$, but it remains
finite in the limit of global load transfer.
Increasing $R$ gives rise to increasing 
slope of $Y(\sigma)$ at the failure point, 
and in the limit of infinite range interaction the tangent of  $Y(\sigma)$ 
becomes vertical at the point of failure.

\subsection{Cluster analysis}
One of the most interesting aspects of the damage mechanism of
disordered solids is that the breakdown 
is preceded by an intensive precursor activity in the form of avalanches
of microscopic breaking events. 
Under a given external load $F$ a certain fraction of fibers 
fails immediately. Due to the load transfer from broken to intact fibers
this primary fiber breaking may initiate secondary breaking that may
also trigger a whole avalanche of breaking. If $F$ is large enough the
avalanche does not stop and the material fails catastrophically.
It has been shown by analytic means that in the case of global load
transfer the size distribution of
avalanches follows asymptotically a universal power law with an
exponent $-5/2$ \cite{hansen3,hansen1}, however, in the case of local
load transfer no universal behavior exists, and the avalanche
characteristic size is bounded 
\cite{hansen2,hansen3}.  This precursory activity can also be 
observed experimentally by means of the acoustic emission analysis.
Acoustic emission measurements have revealed that for a broad
variety of disordered materials the
response to an increasing external load takes place in
bursts having power law size distribution over a wide range
\cite{ciliberto,strauven,ae}.

\begin{figure}[htb]
\begin{center}
\epsfig{bbllx=161,bblly=257,bburx=455,bbury=556,
file=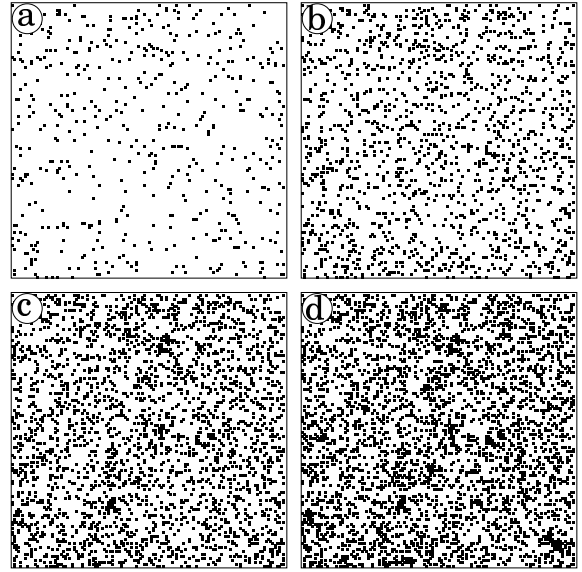,
  width=8cm}
\caption{Snapshots of the damage in the dry FBM model on a square lattice 
of size $L = 128$ 
for different values of the load: a) $F/F_c=0.153$
b) $F/F_c= 0.468$, c) $F/F_c=0.796$ d) $F/F_c=0.997$.}
\label{fig:8}
\end{center}
\end{figure}

In this section we analyze the evolution of damage in local
load transfer FBM, comparing dry and continuous damage models. Instead
of avalanches of fiber failures, we focus on the properties of
clusters of broken fibers which are much less explored.  
In the following simulations we employ a uniform
distribution for the thresholds $d_i$.  
The load after fiber failure is redistributed on the surviving 
nearest neighbors.
As the load is increased, fiber breaks and clusters of broken fibers 
are formed due to the
spatial correlation introduced by the local load transfer.
These clusters of broken fibers can be identified as microcracks formed 
in the plane perpendicular to the load direction.
We monitor the damage evolution by taking snapshots of the clusters
at different loads. In Fig.~\ref{fig:8} the damage evolution is shown
in the dry FBM. One can observe the nucleation and gradual growth of
clusters with increasing load $F$.
We find that the clusters are small compared to the system size even
before global failure (see Fig.\ \ref{fig:8}(d)), 
in accordance with the first-order transition scenario. 
In the continuous damage FBM with local load transfer the clusters of 
failed fibers are defined as connected sets of fibers having the same 
number of failure, taking into account only nearest neighbor 
connections. In these calculations  we set $k_{max}=\infty$, and the 
simulations are stopped when the plastic regime is reached.

\begin{figure}[htb]
\begin{center}
\epsfig{bbllx=125,bblly=375,bburx=467,bbury=680,
file=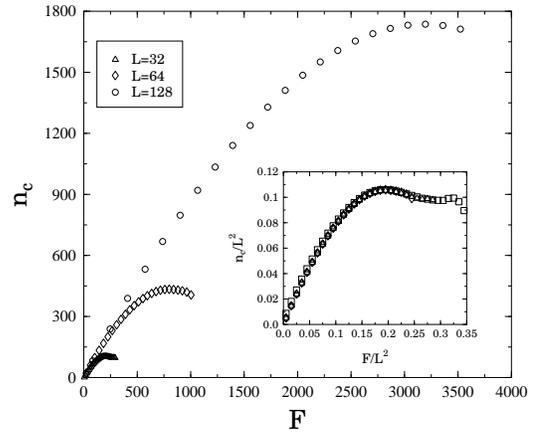,
  width=7cm}
\caption{The number of clusters as a function of the load $F$
in the local stress transfer dry FBM for different system sizes.
In the inset the rescaled plot is presented, where also the system
$L=16$ is shown (square).}
\label{fig:9}
\end{center}
\end{figure}

\begin{figure}[htb]
\begin{center}
\epsfig{bbllx=125,bblly=375,bburx=467,bbury=680,
file=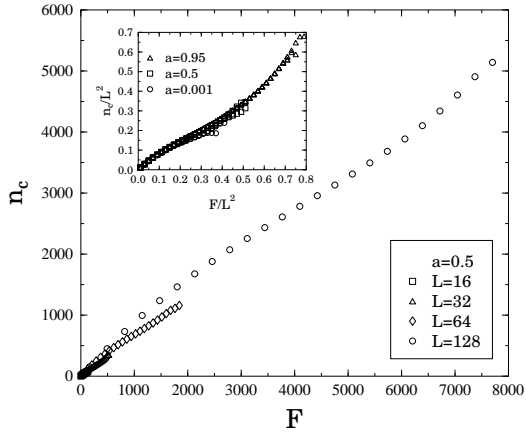, width=7cm}
\caption{The number of clusters as a function of the load $F$ 
in the local stress transfer continuous damage annealed
FBM for different system sizes at $a=0.5$.
In the inset we show the rescaled plot for different values
of $a$ and $L$.}
\label{fig:10}
\end{center}
\end{figure}

To obtain quantitative informations on the damage evolution,
we measure the cluster probability distribution $n(s,F)$, defined as
the number of clusters formed by $s$ neighboring broken fibers when the
applied load is $F$ \cite{zrsv}. 
The moments ($M_k(F) \equiv \int s^k n(s,F) ds$
is the $k$-th moment) of $n(s,F)$ contain most of the information
on the evolution of the damage. 
We determine $n(s,F)$ for different system sizes $L$ by
averaging over the disorder. 
The total number of clusters $n_c\equiv M_0$ as a function of the load 
is presented in Fig.~\ref{fig:9}. 
The increasing part of $n_c$ as a function of $F$ is due to the nucleation 
of new microcracks, and the short plateau or decreasing regime in the vicinity
of global failure is caused by the coalescence of growing cracks.
The inset of Fig.~\ref{fig:9} demonstrates 
that in dry FBM $n_c$ obeys a simple scaling
law $n_c = L^2 g(F/L^2)$ implying that the clusters are homogeneously
scattered through the lattice. A similar scaling is observed for the
continuous damage annealed FBM in Fig.~\ref{fig:10}. The inset shows that 
the scaling function is independent of the damage
parameter $a$. 

\begin{figure}[htb]
\begin{center}
\epsfig{bbllx=140,bblly=60,bburx=470,bbury=680,
file=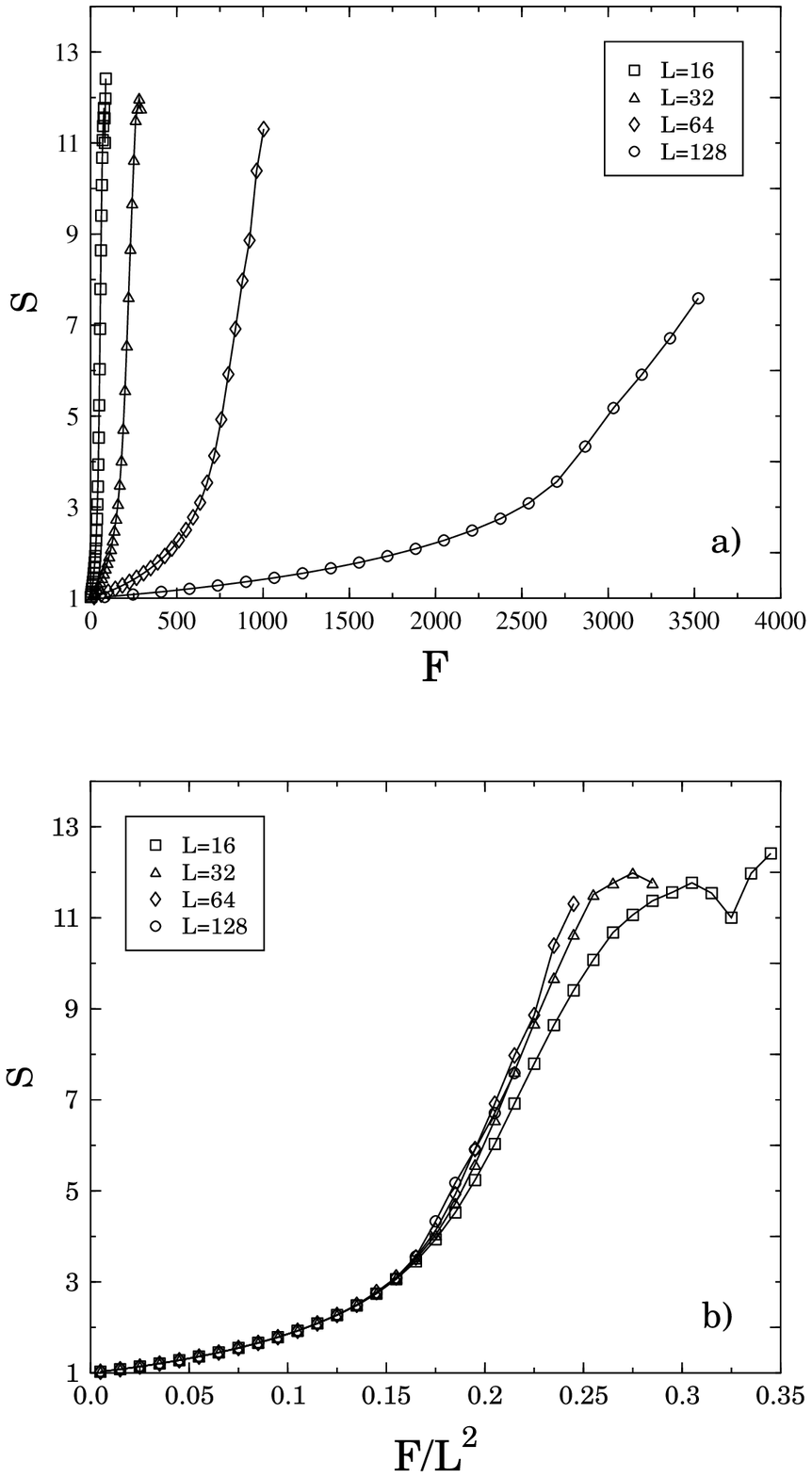,
  width=7cm}
\caption{(a) The average cluster size as a function of the load $F$ 
in the local stress transfer dry FBM for different system sizes $L$
and the corresponding rescaled plot (b).}
\label{fig:11}
\end{center}
\end{figure}

\begin{figure}[htb]
\begin{center}
\epsfig{bbllx=140,bblly=60,bburx=470,bbury=680,
file=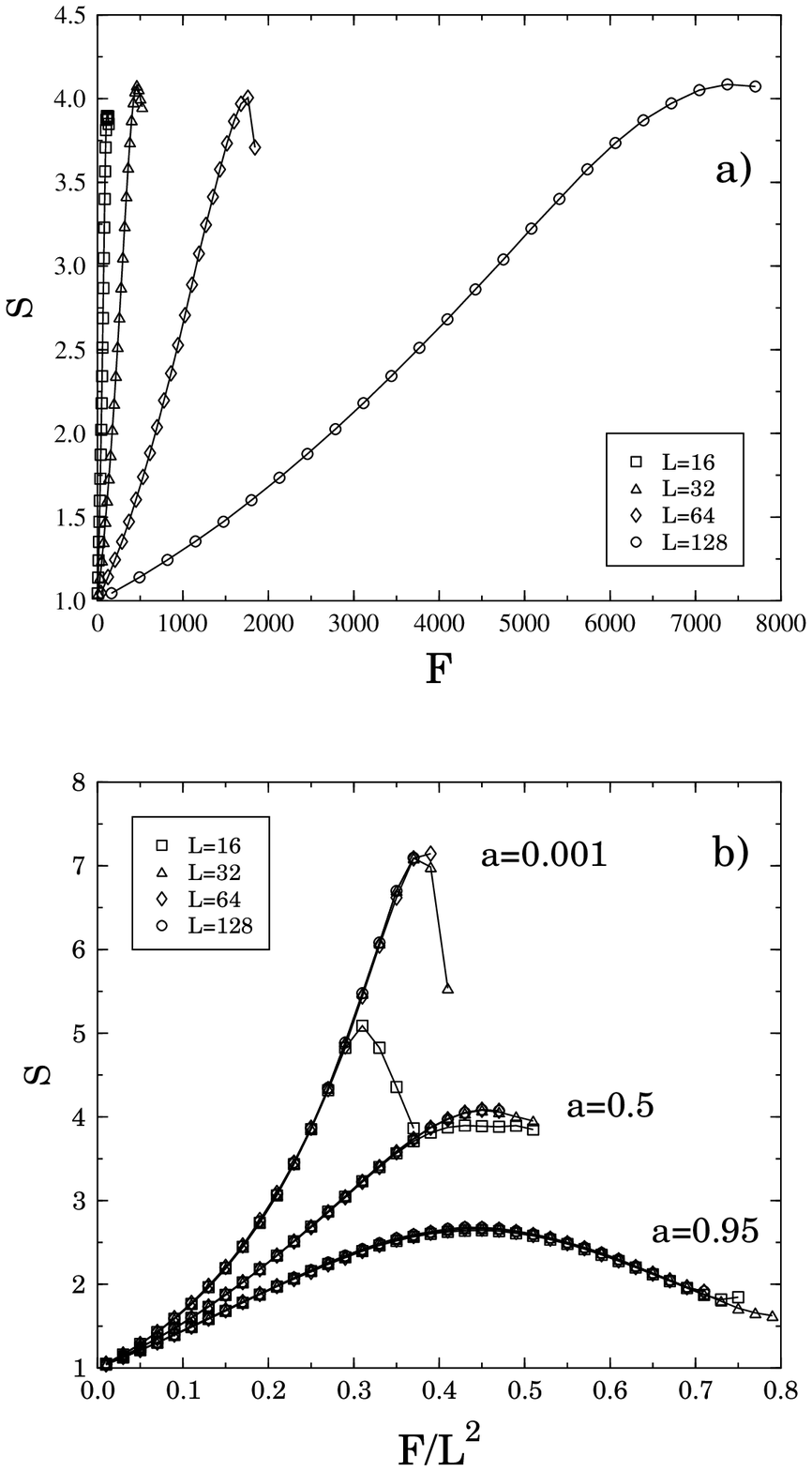, width=7.0cm}
\caption{(a)The average cluster size as a function of the load $F$
in the local stress transfer continuous damage annealed
FBM with $a=0.5$ for different system sizes $L$, 
and (b) the corresponding rescaled plot for different values
of $L$ and $a$.}
\label{fig:12}
\end{center}
\end{figure}

\begin{figure}[htb]
\begin{center}
\epsfig{bbllx=125,bblly=375,bburx=467,bbury=680,
file=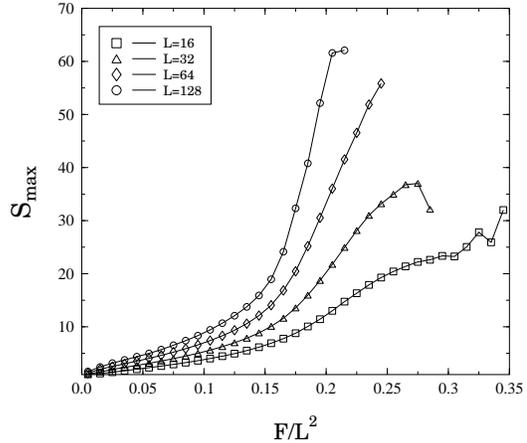, width=7cm}
\caption{The size of the largest cluster $S_{max}$ 
as a function of  $F/L^2$ in the local stress transfer dry
FBM for different system sizes $L$.}
\label{fig:13}
\end{center}
\end{figure}

Next, we measure the average cluster size defined as $S\equiv M_2/M_1$
and show that it approaches a value which decreases with system size
(Fig.~\ref{fig:11}(a)). It can be seen that for a given system size the
$S(F)$ curves have two regimes: a slowly increasing initial regime due to
the nucleation and growth of clusters, and a rapidly increasing part
close to global failure which is caused by the coalescence of growing
cracks. 
In dry FBM, we can simply rescale the data according to
the law $S(F,L)=s(F/L^2)$ and obtain a good collapse (Fig.~\ref{fig:11}(b)).
Similar results are obtained for the continuous damage case, but 
we see that the rescaled curves depend on $a$ (Fig.~\ref{fig:12}). 
The larger $a$ is, the smaller the 
clusters are, since the stress concentration decreases with increasing 
$a$, and the disorder gets more dominating. 
These results demonstrate that global failure is initiated once the
crack size reaches a critical size $s_c$ after which a crack
becomes unstable. However, the average cluster size $S$ does not
provide a reliable estimate of $s_c$, which can be obtained instead 
monitoring the size of the largest cluster $S_{max}$ as a function of the 
load. It can be seen in Fig.\ \ref{fig:13} that $S_{max}$ reaches a value
that increases with $L$, 
but for large $L$ this value seems to saturate. The rapid increase 
of $S_{max}$ close to the failure point is due to the coalescence of
clusters and is thus produced by a very small amount of fiber failures.

\section{damage localization: effect of the quenched disorder}

In the previous section, we analyzed the damage
structure in the local load transfer  models. In the dry 
FBM and in the case of continuous damage FBM with annealed disorder we
do not expect to find any non trivial damage localization for global 
load transfer rules, since these models behave effectively
like in mean-field theory. On the other hand, quenched disorder
can lead to localized structure for continuous damage 
FBM even in global load transfer conditions. Weak fibers
are expected to fail more times generating an inhomogeneous
damage pattern.  

\begin{figure}[htb]
\begin{center}
\epsfig{bbllx=140,bblly=60,bburx=470,bbury=680,
file=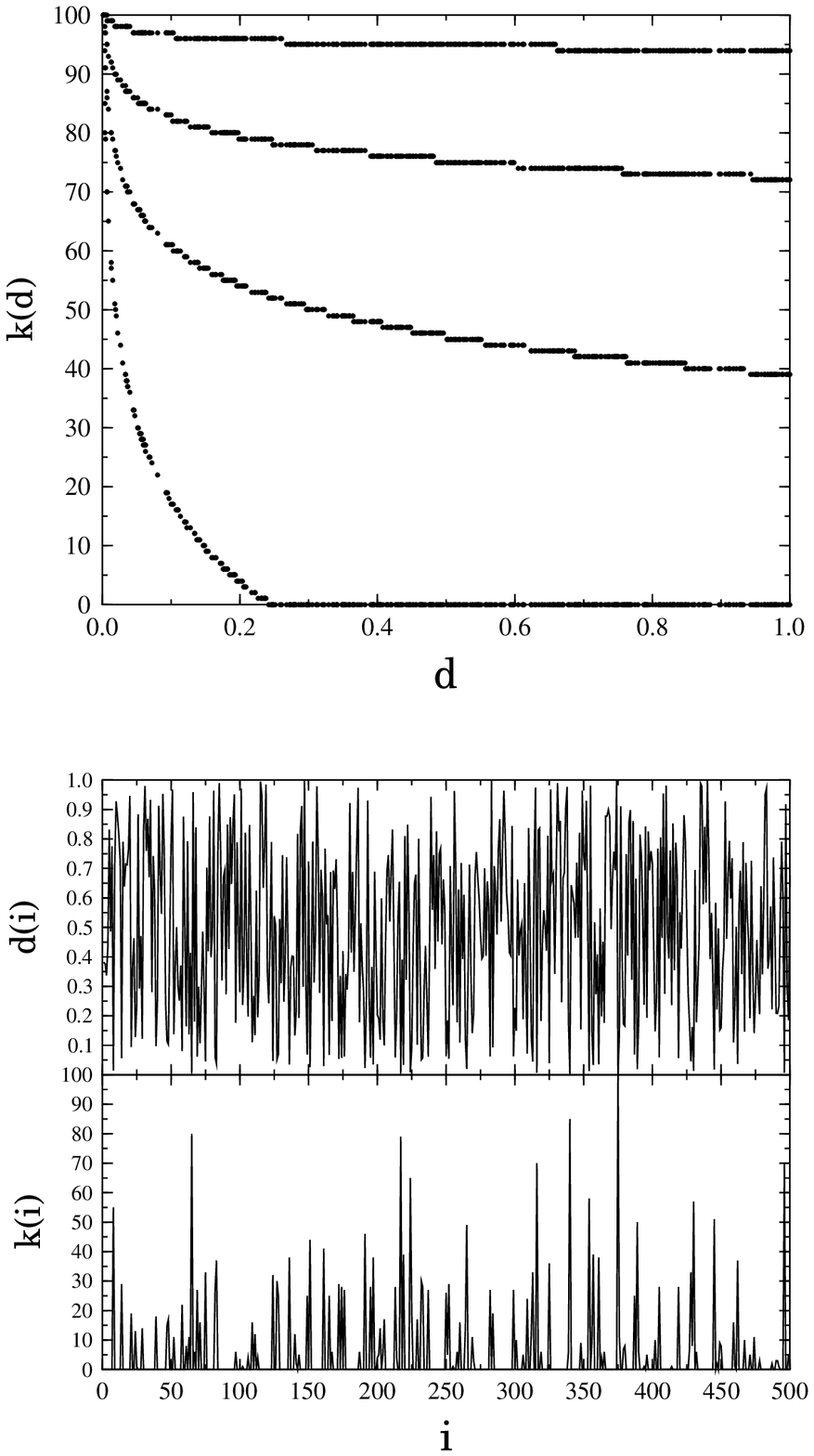, width=7.0cm}
\caption{The number of failures $k$ of the fibers as a function of
their threshold $d$ (distributed uniformly)
in the quenched global load transfer continuous FBM (top). The
parameters used are $k_{max}=100$, $N=500$ and $a=0.4, 0.8, 0.9,  
0.95$.  The threshold $d(i)$ as a function of $i$
is compared with the 'damage' $k(i)$ for $a=0.9$  (bottom). } 
\label{fig:14}
\end{center}
\end{figure}

\begin{figure}[!ht]
\begin{center}
\epsfig{bbllx=125,bblly=330,bburx=467,bbury=640,
file=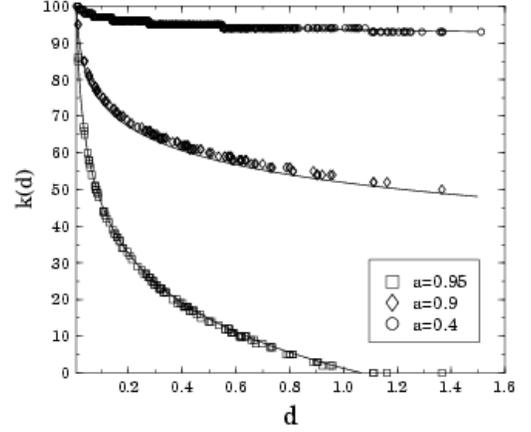, width=7cm}
\caption{The number of failures $k$ of the fibers as a function of
their threshold $d$ (distributed according to the Weibull distribution
with $m=1.5$) in the quenched global load transfer continuous
FBM: comparison between simulations and analytic results
for $a=0.95, 0.9, 0.4$.}
\label{fig:15}
\end{center}
\end{figure}

In order to analyze this effect, we measure
$k(i)$, the number of failures at fiber $i$, when there is exactly one
fiber, which has reached $k_{max}$. In the bottom part of 
Fig.~\ref{fig:14} we plot the value of 
$k(i)$ and the corresponding value of the threshold $d(i)$.
The 'damage' $k(i)$ shows a very irregular pattern, which
should be compared with a roughly uniform structure expected
for annealed disorder. In the upper part of the figure we
display the decay of $k$ as a function of $d$, showing how
weak fibers break more often than strong ones. The decay
is  more pronounced when $a$  is close to one, and 
becomes less important for smaller $a$.
It is straightforward to obtain an analytic expression
for $k(d)$ which is compared with the numerical results
in Fig.~\ref{fig:15}.

\begin{figure}[htb]
\begin{center}
\epsfig{bbllx=125,bblly=375,bburx=467,bbury=680,
file=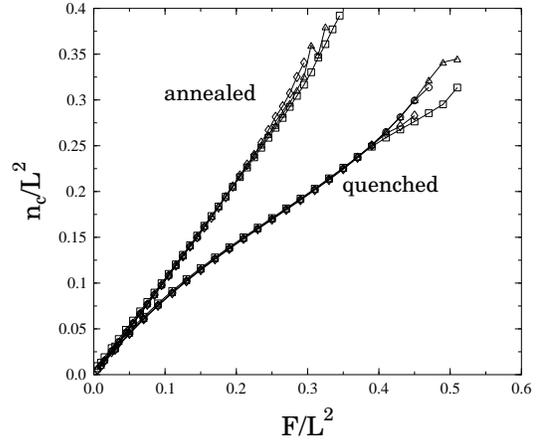, width=7cm}
\caption{The number of clusters as a function of the rescaled
load $F/L^2$  in the local stress transfer continuous damage quenched and
annealed FBM for different system sizes at $a=0.5$. Note that
the number of clusters increses faster for annealed disorder
indicating a smaller degree localization.}
\label{fig:16}
\end{center}
\end{figure}

\begin{figure}[htb]
\begin{center}
\epsfig{bbllx=125,bblly=375,bburx=467,bbury=680,
file=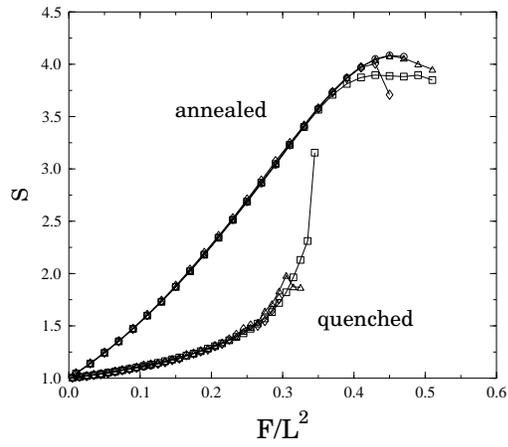, width=7cm}
\caption{The average cluster size as a function of the rescaled
load $F/L^2$  in the local stress transfer continuous damage quenched and
annealed FBM for different system sizes at $a=0.5$. Note that
the cluster size increses faster for quenched disorder
indicating a larger degree localization.}
\label{fig:17}
\end{center}
\end{figure}

Finally, we expect that quenched disorder should have
an effect also on the cluster structure of local load
transfer models. In order to confirm this point, we
compare the number of clusters $n_c(F)$ and the 
average cluster size $S$ for quenched and annealed disorder.
The results, shown in Figs.~\ref{fig:16} and \ref{fig:17},
indicate that damage is more localized when the disorder
is quenched.

\section{conclusions}

We have proposed a continuous damage
version of the FBM, that can be used to model a wide variety of 
constitutive behaviors. 
We have analyzed the development of damage in FBM under different
conditions and compared the local load transfer model with 
different interaction ranges with the global load transfer model
that can be solved exactly in the limit $N\to\infty$. 

From the theoretical point of view, the cluster analysis 
shows the analogies between the failure  in the  FBM and 
nucleation in first-order phase transition. The failure point
in global load transfer FBM plays the role of a spinodal point.
The strain $f$ carried by the fibers, proportional to the
fraction of intact fibers \cite{zrsv}, close to the failure load 
$F_c$ has a diverging derivative $df/dF \sim (F_c-F)^{-1/2}$.
A similar behavior is observed close to a spinodal instability
in first-order phase transitions. One should note that the spinodal
point and its associated scaling is a mean-field property, obtained
in the limit $N\to\infty$, and 
in general is not observed for finite dimensional short-ranged 
models where nucleation occurs much before reaching the spinodal.
Similarly, in local-load transfer (i.e. short range) FBM  
failure occurs much before the corresponding global
load sharing instability (i.e. the spinodal) and
the avalanche characteristic size is bounded 
\cite{hansen1,hansen2,hansen3}.

We have shown that increasing the range of interaction 
the failure point is shifted towards the spinodal. A similar
behavior is observed for instance in Ising systems when the
range of interactions is increased \cite{raykl}.
In models with long range stress transfer, as for instance
in elastic or electric networks, it is possible to observe the
spinodal scaling even in finite dimensional systems 
\cite{zrsv,hansen3}. The presence of a spinodal instability 
could explain the observation of scaling properties in acoustic
emission experiments \cite{ciliberto,strauven,ae}.

Our continuous damage FBM can reproduce 
a wide variety of elasto-plastic constitutive behaviors.
A remarkable feature of the model is that multiple failure 
of the elements results in ductile macroscopic behavior 
in spite of the brittleness of the constituents. 
Similar ductile behavior has been observed experimentally 
in fiber reinforced composites made of brittle constituents
\cite{psh,pshtheory2,pshtheory3}.  
Experiments revealed that the mechanism of this ductility,
which is called pseudo-strain hardening, is the multiple failure
of the material \cite{psh,pshtheory2,pshtheory3}. 
Our continuous damage model, recovering 
as special cases the dry bundle model and micromechanical models 
known in the literature, could provide a general framework for the 
statistical-micromechanical modeling of the  behavior of fiber 
reinforced composites. The fitting of experimental results in the
framework of our model will be presented in a forthcoming
publication.

\section*{Acknowledgment}
This work was supported by the project SFB381.
F.\ K.\  acknowledges financial support from 
the Alexander von Humboldt Foundation
(Roman Herzog Fellowship). F.\ K.\ is grateful to I.\ Sajtos for the
valuable discussions.
S. Z. acknowledges 
financial support from EC TMR Research Network
under contract ERBFMRXCT960062.

\end{document}